\documentstyle[12pt]{article}
 
\def\a{\alpha}
\def\b{\beta}
\def\d{\delta}      \def\D{\Delta}

\def\g{\gamma}      \def\G{\Gamma}
\def\h{\eta}

\def\l{\lambda}

\def\o{\omega}   
\def\p{\pi}      
   
\def\r{\rho}

\def\bo{{\raise.15ex\hbox{\large$\Box$}}}

\def\dag{^{\dagger}{}}
\def\rarr{\rightarrow}

\def\ordless{{\lower2mm\hbox{$\,\stackrel{\textstyle <}{\sim}\, $}}}
\def\ordmore{{\lower2mm\hbox{$\,\stackrel{\textstyle >}{\sim}\, $}}}

\newtoks\slashfraction
\slashfraction={.13}
\def\slash#1{\setbox0\hbox{$\, #1$}
\setbox0\hbox to \the\slashfraction\wd0{\hss \box0}/\box0}

\def\leftrightarrowfill{$\mathsurround=0pt \mathord\leftarrow \mkern-6mu
        \cleaders\hbox{$\mkern-2mu \mathord- \mkern-2mu$}\hfill
        \mkern-6mu \mathord\rightarrow$}
\def\overleftrightarrow#1{\vbox{\ialign{##\crcr
        \leftrightarrowfill\crcr\noalign{\kern-1pt\nointerlineskip}
        $\hfil\displaystyle{#1}\hfil$\crcr}}}
\def\startarray{\left( \begin{array}}
\def\finarray{\end{array} \right)}
\def\starteq{%\vspace{.1in}
\begin{eqnarray}}
\def\fineq{\end{eqnarray}
}

\catcode`@=11
\def\underline#1{\relax\ifmmode\@@underline#1\else
$\@@underline{\hbox{#1}}$\relax\fi}
\catcode`@=12
 
\newskip\humongous \humongous=0pt plus 1000pt minus 1000pt

\newif\ifdtup
 
\def\NP#1{{\em Nucl. Phys.} {\bf B{#1}}}
\def\PL#1{{\em Phys. Lett.\ }{\bf B{#1}}}

\def\PRD#1{{\em Phys. Rev.\ }{\bf D{#1}}}

\def\zph#1{{\em Z.~Phys.\ }{\bf C{#1}}}

\def\textcite#1{Ref.~{\cite{#1}}}
\def\Ref#1{(\ref{#1})}
\def\Eqn#1{Eq. \Ref{#1}}

\def\thefootnote{\fnsymbol{footnote}}
 
\def\author#1#2{{\bf #1} \\ {\em #2}\vspace{5mm}}
\def\ecolepoly{Centre de Physique Th\'eorique, \\
Centre National de la Recherche Scientifique, UPR A0014, \\  
Ecole Polytechnique, 91128 Palaiseau Cedex, France}

\def\bold#1{\setbox0=\hbox{$#1$}%
     \kern-.025em\copy0\kern-\wd0
     \kern.05em\copy0\kern-\wd0
     \kern-.025em\raise.0433em\box0 }

\def\title#1#2#3#4#5{\thispagestyle{empty}
        \begin{center} \vspace*{1cm} { \bf #3} \\[.5in] {#4{}}
        \end{center} \vfill \centerline{ ABSTRACT}
   {\nopagebreak \noindent\begin{quotation}\noindent {\small #5}
   \end{quotation}} \vfill {#2} \hfill\begin{tabular}{r} {#1} 
        \end{tabular}  \newpage
        \def\thefootnote{\arabic{footnote}}}
\def\prefer{\section*{}
    \list{[\arabic{enumi}]}{\usecounter{enumi}\settowidth\labelwidth{[000]}
      \leftmargin\labelwidth\advance\leftmargin\labelsep \rightmargin=0pt}
        \small \sfcode`\.=1000\relax}

\def\refer#1{\section*{\large \sc {#1}}
    \list{\arabic{enumi}.}{\usecounter{enumi}\settowidth\labelwidth{[000]}
      \leftmargin\labelwidth\advance\leftmargin\labelsep \rightmargin=0pt}
        \raggedright \small \sfcode`\.=1000\relax}

\def\ReFer#1#2{\section*{\large\sc#1}
    \list{[\arabic{enumi}]}{\usecounter{enumi}\settowidth\labelwidth{#2}
      \leftmargin\labelwidth\advance\leftmargin\labelsep \rightmargin=0pt}
        \raggedright \small \sfcode`\.=1000\relax}

\def\REFER#1#2{\section*{\large\sc#1}
    \list{#2 {enumi}.}{\usecounter{enumi}\settowidth\labelwidth{[000]}
      \leftmargin\labelwidth\advance\leftmargin\labelsep \rightmargin=0pt}
        \raggedright \small \sfcode`\.=1000\relax}

\def\startbib{\vspace{1in}\begin{refer}{References}
\small\frenchspacing\nopagebreak}
\def\endbib{\end{refer} \normalsize \nonfrenchspacing}
\def\startfig{\newpage \centerline{{\sl Figure captions}} \begin{itemize}}

\def\endfig{\end{itemize}}
 
\begin{document}
\title {September 1996} {CPT-S466.0996/BARI-TH/96-253} {FINAL STATE INTERACTIONS IN 
$B \rarr D \r$  and $B \rarr D^{*}\p$ DECAYS 
}
 {\author { G. Nardulli$^{(*)}$ \footnotetext{(*) e-mail:nardulli@ba.infn.it}}
 {Dipartimento di Fisica, Universit\`a di Bari, Italy and \\
INFN,Sezione di Bari, Italy} \\
 \author { T. N. Pham${\dag}$ \footnotetext{${\dag}$  e-mail:pham@orphee.polytechnique.fr} }  {\ecolepoly}}
{We analyze final state strong interaction effects in 
 $B \rarr D \r$ and $B \rarr D^{*}\p$ decays using the Regge model. We
find that, due to
the smallness of the contributions from  the non-leading Regge trajectories
($\r$ , $f$, $\p$ etc.), final state interaction phases are small if
the Pomeron coupling to the charm quark is suppressed
in comparison to lighter quarks. Our conclusion is
that for $B$ decays into states containing
charm, final state interaction effects should play a minor role.}

  The problem of final state strong interactions in non-leptonic heavy
meson decays has recently received considerable theoretical 
interest \cite{Buccella,Lusignoli,Despande,Zheng,Donoghue}. 
The relevance of the problem is
related to the present and future programs for studying $CP$-violation
in heavy hadrons, in particular $B$ meson, decays. As well known \cite{CP},
$CP$-violation in such systems might be observed by measuring an 
interference effect between two different amplitudes, and the relevant
physical observable, i.e the $CP$-odd asymmetry, turns out to depend
crucially on the strong interaction phase difference between the two
amplitudes.

  Whereas the usual approach for charmed meson decays is the
parametrization of the final state interaction effects by means of a 
resonant rescattering of the
final particles, for $B$ decays
final state interaction effects are in general expected to be 
small. This seems rather plausible because the final
decay products are moving away from each other with large momentum;
due to the relativistic time dilatation, the formation time
of the final particles is large and, when formed, they are far away
from the color sources, which implies that strong phases 
induced by the color interactions should be small \cite{Bj} (see also
\cite{Neubert}). Some evidence for this comes for example from
exclusive non-leptonic two-body $B$ decays \cite{Kamal}.

  The expectation that the rescattering effects in the final state due to
soft interactions become negligible in the $m_{B} \rarr \infty$ limit
has been challenged in \cite{Donoghue}. Taking $B \rarr \p\p$ and
$B \rarr K\p$ as examples, these authors assume Pomeron dominance 
and Regge theory \cite{Collins}
at high energy (i.e. large $\sqrt{s}= m_{B}$) to estimate the size of the
inelastic final state interaction and its effect on $CP$-violating
asymmetries, such as 
$\D \G = \G(B^{-} \rarr K^{-}\p^{0}) - \G(B^{+} \rarr K^{+}\p^{0}) $;
their conclusion is that these inelastic effects, mainly induced by
the Pomeron,  are sizeable. Since a
reliable way to compute these soft, non perturbative, effects is
missing at the moment, the rather pessimistic conclusion reached 
in \cite{Donoghue} is that the presence of final state interactions
will limit the accuracy of the standard approaches to $B$ decays 
that are based on the use of perturbative QCD
supplemented by the factorization hypothesis, even though
in the  $m_{B} \rarr \infty$ limit
both  methods are expected to be more and more reliable.

 A rather different conclusion has been reached, on the other hand, by
Zheng in \cite{Zheng}. Using basically the same approach (i.e. an
approximate evaluation of the final state strong interaction $S$-matrix
based on the Regge model) and considering $B \rarr D K$ non leptonic
decays, Zheng concludes that the approximation of neglecting final state
strong phases is `` accurate up to, roughly speaking, about $10 \%$''.
$ $
  Because of these conflicting results, we believe worthwhile to
investigate more accurately this problem. Therefore, in the present
paper, we consider other decay channels: $B \rarr D \r,D^{*}\p $ using
the same model (Regge theory) already employed in \cite{Zheng} 
and \cite{Donoghue}. Apart from the choice of a different decay mode,
our treatment differs from \cite{Zheng} and \cite{Donoghue} because 
we explicitly compute some inelastic effects, i.e. 
the  $D \r \rarr D^{*}\p $ rescattering. The two-body $D^{*}\pi$,
$D \rho$ $B$ decays are presently investigated by several
experiments (for a review see \cite{hon}) and present
a noticeable theoretical  and experimental interest, related to the validity
of the factorization approximation and to the sign of the ratio of Wilson
coefficients $a_2/a_1$.

  For decay amplitudes, final state interactions are taken into
account by means of the Watson's theorem \cite{Neubert,Watson}:

\starteq
A = \sqrt{S}A_{b}
\label{amp} 
\fineq
where $S$ is the $S$-matrix, $A_{b}$ are the 
bare amplitudes, i.e. decay amplitudes with no final state interactions,
and $A$ are the full amplitudes. The $S$-matrix relates amplitudes
with the same isospin $I$ and a given total angular momentum $J$. Since
we consider the amplitudes $A^{I}_{1}= A(B \rarr D^{*}\p)_{I}$ 
and $A^{I}_{2}= A(B \rarr D \r)_{I}$, the final state has $J=0$;
moreover the vector mesons ($\r$ or $D^{*}$) in the final state can
only have longitudinal helicity ($\l = 0$), which means that,
effectively, we are dealing with a situation analogous to a decay into
scalar particles, which simplifies considerably the formalism.

  Let us begin by writing down explicitly the isospin amplitudes ($I=3/2$ 
and $I=1/2$) in terms of the physical amplitudes

\starteq 
A(B \rarr D^{*}\p)_{3/2} & =& - \sqrt{2/3}
A(B^{0} \rarr \bar{D}^{*0}\p^{0}) + 
\sqrt{1/3}A(B^{0} \rarr D^{*-}\p^{+}) \  \nonumber \\
A(B \rarr D^{*}\p)_{1/2} & =& \sqrt{1/3}
A(B^{0} \rarr \bar{D}^{*0}\p^{0}) + 
\sqrt{2/3}A(B^{0} \rarr D^{*-}\p^{+}) 
\label{isoamp1} 
\fineq
\par\noindent
and, similarly~:
\starteq 
A(B \rarr D\r)_{3/2} & =& - \sqrt{2/3}
A(B^{0} \rarr \bar{D}^{0}\r^{0}) + 
\sqrt{1/3}A(B^{0} \rarr D^{-}\r^{+}) \  \nonumber \\
A(B \rarr D\r)_{1/2} & =& \sqrt{1/3}
A(B^{0} \rarr \bar{D}^{0}\r^{0}) + 
\sqrt{2/3}A(B^{0} \rarr D^{-}\rho^{+}) ~~. 
\label{isoamp2} 
\fineq

The $J=0$ $S$-matrix, for each given isospin channel, must satisfy the
unitarity relation. In standard notation, the two-body $S$-matrix
elements are given by ($i,j=1,2$)
\starteq 
S^{(0)I}_{ij} = \d_{ij} + 2i\sqrt{\r_{i}\r_{j}}A^{(0)I}_{ij}~~,
\label{s-matrix} 
\fineq
\par\noindent
where the $J=0$, isospin $I$ amplitude $A^{(0)I}_{ij}= 
A^{(0)I}_{ij}(s)$ is obtained by
projecting the $J=0$ angular momentum out of the amplitude $A^{I}_{ij}(s,t)$:
\starteq 
A^{(0)I}_{ij}(s) =  {1 \over 16\p}{s \over \sqrt{\l_{i}\l_{j}}}
\int^{t_{-}}_{t_{+}}dt \,A^{I}_{ij}(s,t) ~~.
\label{S-wave} 
\fineq

Working in the approximation $m_{D}= m_{D^{*}}$, 
$m_{\p}\simeq m_{\r}\simeq 0$, with $\sqrt{s}= m_{B}$, we have
$\l_{i}= (s - m_{D}^{2})^{2}$, 
$\r_{i}={\sqrt{ \l_{i}}}/s \simeq (s - m_{D}^{2})/s$, $t_{-}=0$ and 
$t_{+}= -(s - m_{D}^{2})^{2}/s$.

 In order to compute \Eqn{s-matrix} and \Eqn{S-wave}, we need 
$A^{I}_{ij}(s,t)$. As we stressed already, in
order to evaluate these amplitudes, we will  work
in the Regge model,  which should be a
reasonable theoretical framework due   to the rather large value of 
$s = m_{B}^{2}$. In terms of the Pomeron ($P$), which is the leading
contribution, and the non-leading trajectories $\r$, $f(1270)$ and $\p$,
neglecting Regge cuts \footnote{ A discussion on Regge cuts is contained 
in \cite{Zheng}.}, we have:
\starteq
A^{3/2}_{11} & =& 
A(\bar{D}^{*}\p \rarr \bar{D}^{*}\p)_{3/2} =  P + f + \r \ \nonumber \\
A^{3/2}_{12} & =& A^{3/2}_{21}  = 
A(D^{*}\p \rarr D \r)_{3/2}  =  \p    \      \nonumber \\
A^{3/2}_{22} & =& 
A(D \r \rarr D \r)_{3/2}  =  P' + f' + \r'
\label{amp3} 
\fineq
and 
\starteq
A^{1/2}_{11} & =& 
A(D^{*}\p \rarr D^{*}\p)_{1/2} =  P + f - 2 \r \nonumber \\
A^{1/2}_{12} & =& A^{1/2}_{21}  = 
A(D^{*}\p \rarr D \r)_{1/2}  =  -2\p    \nonumber \\
A^{1/2}_{22} & =& 
A(D \r \rarr D \r)_{1/2}  =  P' + f' -2 \r'~~.
\label{amp1} 
\fineq

 We observe that the primed ($P'$, $f'$, $\r'$) contributions may
differ from the unprimed ones only for a numerical coefficient; we also
observe that the leading Regge trajectories in the off-diagonal matrix
elements should be $\o$ and $A_{2}$; however they only contribute to the
helicity-flip amplitudes (with $\l = \pm 1$) that are not of interest 
here, which is why we take into account the next-to-leading
trajectory, i.e. the pion ($\p$) Regge exchange. 

Let us first consider
the Pomeron contribution, that we parametrize as follows
\starteq
P =~-~ \b^{P}g(t) \left({s \over s_{0}}   \right)^{\a_{P(t)}}
e^{ -i{\p \over 2}\a_{P}(t)}~~,
\label{P} 
\fineq
with $s_{0} = 1\,\rm GeV^{2}$ and 
\starteq
\a_{P}(t) = 1.08 + 0.25t   \qquad \qquad (t {\rm \ in \ GeV^{2}})~~,
\label{alphaP} 
\fineq
as given by fits to hadron-hadron scattering total cross 
sections \cite{Donnachie}, \cite{PDG}. The product
$\b^{P}\cdot g(t) = \b^{P}(t)$ represents the Pomeron residue; for the
$t$-dependence we assume
\starteq
g(t) = {1 \over (s- t/m_{\r}^{2})^{2}} \simeq e^{2.8t}~~,
\label{gt} 
\fineq
which is motivated by the analogy with the electromagnetic form factor
and by the smallness of $t$, due to the exponential damping in
$(s/s_{0})^{\a_{P}(t)} $. As for the residue at $t=0$, i.e. $\b^{P}$, we
assume, as usual, factorization:
\starteq
\b^{P} = \b^{P}_{D^{*}}\b^{P}_{\p}
\label{betaP} 
\fineq
for the elastic $D^{*}\p \rarr D^{*}\p$ amplitude.

  The residue at the vertex $P\p\p$ can be extracted from proton-proton
and pion-proton high energy scattering; we find
\starteq
\b^{P}_{\p} \simeq {2 \over 3}\b^{P}_{p} = 5.1
\label{betaPpi} 
\fineq
which is consistent with the hypothesis of the additive quark counting
rule: $\b^{P}_{\p}= 2\b^{P}(uu)$,  $~~ \b^{P}_{p}= 3\b^{P}(uu)$.

  It is worthwhile to remark at this stage that we can obtain, from
$\g p$ high energy scattering data, the $P\r\r$ residue $\b^{P}_{\r} $ 
by making the assumption of Vector Meson Dominance (VMD); in this
way we find the approximate relation
\starteq
\b^{P}_{\r} \simeq \b^{P}_{\p}
\label{betaPrho} 
\fineq
which is numerically valid 
within $15\%$. \Eqn{betaPrho} is also consistent with
the additive quark counting rule. As for the coupling of the Pomeron to
charm, assuming again the additive quark model, one has
\starteq
\b^{P}_{D} = \b^{P}_{D^{*}} = \b^{P}(cu) + \b^{P}(uu)
\label{betaPD} 
\fineq
and for $\b^{P}(cu) $ one has to assume as an input some theoretical
ansatz; for example in \cite{Zheng} it is assumed:
\starteq
 \b^{P}(cu) = {1 \over 10} \b^{P}(uu)~.
\label{betaPcu} 
\fineq
  We shall assume \Eqn{betaPcu} as well and will comment on this choice
below. Let us observe that, by this assumption, we find
\starteq
P = P' \quad ,
\label{P2} 
\fineq
a result to be used in \Eqn{amp3}-\Eqn{amp1}.

  Let us now consider the non-leading Regge trajectories 
$R (= \r,f,\p)$. For these exchanges we write the general formula
\starteq
R = -\b^{R} {1 + (-)^{s_{R}}
e^{ -i\p\a_{R}(t)} \over 2}\G(l_{R} - \a_{R}(t))(\a ')^{1 - l_{R}}
(\a 's)^{\a _{R}(t)}
\label{R} 
\fineq
as suggested in \cite{Irving}. $\a_{R}(t)$ is the Regge trajectory
given by
\starteq
 \a_{R}(t) = \a_{R}(0) + \a 't \ ,
\label{alphaR} 
\fineq
with an universal slope $\a '= 0.93 \, \rm GeV^{-2}$ and
$\a_{\r}(0)=  \a_{f}(0) = 0.44$ and $\a_{\p}(0)= - \a ' m_{\p}^{2}$.
$l_{R}$ is the lowest spin occurring in the exchange degenerate trajectory
($l_{\r} = l_{f}= 1$, $l_{\p} = 0$) and $s_{R}$ is the spin of the
exchanged meson in the Regge amplitudes 
($s_{\r} = 1 $, $s_{f}= 2$, $s_{\p} = 0$). The choice \Eqn{R} for the
Reggeized amplitudes is suggested by the high energy limit of a
Veneziano amplitude. Since $\a_{R}(t) = s_{R} + \a '(t - m_{R}^{2}) $,
near $t = m_{R}^{2}$, \Eqn{R} reduces to 
\starteq
R \approx \b^{R}{s^{s_{R}} \over (m_{R}^{2} - t)}
\label{Rapprox} 
\fineq 
which allows us to identify $\b^{R} $ as the product of two on-shell
coupling constants (as in \cite{Irving}, we neglect here  the $t$-dependence
 of the Regge residues\footnote{We have verified that this
does not alter our numerical conclusions.}).

  To compute the different residues, we assume factorization, exchange
degeneracy and $SU(4)$ symmetry (which is of course largely violated,
but should at least provide us with an order of magnitude estimate).
Therefore we have $\b^{\r}= \b^{\r}_{\p\p} \b^{\r}_{D^{*}D^{*}}$, and
also $\b^{\r}_{DD}= \b^{\r}_{\p\p} $ and 
$\b^{\r}_{\r\r} =  \b^{\r}_{D^{*}D^{*}}$, Using as an input the 
$g_{\r\p\p}$ coupling constant and Vector Meson Dominance to relate
$\b^{\r}_{\r\r} $ to the electromagnetic coupling of the $\r^{+}$
particle, we finally obtain $\b^{\r} =\b^{f} \simeq 23 $ and $\r = \r'~,
~f=f'$.
As for the pion exchange, we obtain, by this method, 
$\b^{\p}_{\p\r_{0}} = 4.6\,\rm GeV $ from $\r \rarr \p\p$ decay, and
$\b^{\p}_{DD^{*}}= 1.7\,\rm GeV$ from a theoretical estimate of the
$D^{*}D\p$ coupling \cite{Colangelo}, with the result 
$\b^{\p} = 7.8\,\rm GeV^{2} $. Using these results, together with
\Eqn{s-matrix}-\Eqn{amp1}, we obtain the following results for the
$J=0$, isospin $I= 3/2$  and $ 1/2$ $2\times 2$ $S$-matrices:
\starteq
S^{(0)3/2} = \pmatrix{0.76 - 0.06i & -(0.17 + 1.6i)\times 10^{-2} \cr
-(0.17 + 1.6i)\times 10^{-2} & 0.76 - 0.06i }
\label{s3} 
\fineq
and
\starteq
S^{(0)1/2} = \pmatrix{0.71 - 0.02i & +(0.34 + 3.2i)\times 10^{-2} \cr
+(0.34 + 3.2i)\times 10^{-2} & 0.71 - 0.02i }
\label{s1} 
\fineq

 Let us now comment on these results. First of all, as one can see, 
$S^{(0)3/2} $ and $S^{(0)1/2} $
 in \Eqn{s3} and \Eqn{s1}  are not unitary matrices, the reason being
that other inelastic effects, besides the $D \r \rarr D^{*}\p $ final state 
interaction, are present. We shall comment on this point later on; for
the time being we observe that, because of the smallness of the 
off-diagonal matrix elements in \Eqn{s3} and \Eqn{s1}, inelastic
$D \r \rarr D^{*}\p $ scattering is not expected to play a major role
in determining, through Watson's theorem\footnote{ We find \\
$\left(\sqrt{S^{(0)3/2}}   \right)_{11}= 
\left(\sqrt{S^{(0)3/2}}   \right)_{22} = 0.87 - 0.03i$ ; 
$\left(\sqrt{S^{(0)3/2}}   \right)_{12}= 
\left(\sqrt{S^{(0)3/2}}   \right)_{21} =-(0.07 + 0.9i)\times 10^{-2} $ \\
$\left(\sqrt{S^{(0)1/2}}   \right)_{11}= 
\left(\sqrt{S^{(0)1/2}}   \right)_{22} = 0.84 - 0.02i$ ; 
$\left(\sqrt{S^{(0)1/2}}   \right)_{12}= 
\left(\sqrt{S^{(0)1/2}}   \right)_{21} =+(0.16 + 1.9i)\times 10^{-2} $ }
, the full ($B \rarr f$) amplitude. For example, the bare amplitude
$A_{b}(B^{0}\rarr \bar{D}^{0}\r^{0})$ does contribute, through
final state interactions, to the decay amplitude 
$A(B^{0} \rarr \bar{D}^{*0}\p^{0})$, but its contribution to the width does
not exceed a few percent. Moreover, the bare amplitudes
$A_{b}(B^{0} \rarr D^{*-}\p^{+})$  and $A_{b}(B^{0} \rarr D^{-}\r^{+})$,
that, in principle, also contribute to $A(B^{0} \rarr \bar{D}^{*0}\p^{0})$
via final state interactions and can destroy the simple predictions
based on factorization\footnote{In the factorization approximation they
are proportional to the Wilson coefficient $a_{1}$, while
 the bare amplitudes $A_{b}(B^{0}\rarr \bar{D}^{0}\r^{0})$ and
$A_{b}(B^{0} \rarr \bar{D}^{*0}\p^{0})$ depend on the much smaller
coefficient $a_{2}$, see e.g.\cite{Deandrea}.}, give
a negligibly small contribution (of the order $2\%$) to the width.

  As mentioned earlier, the above results are obtained in the Regge
model with exchange degeneracy. One could avoid assuming exchange degeneracy
and take different residues and intercepts for the $\o$ and $f$ as
trajectories fitted by total cross sections and elastic scattering
data \cite{Giacomelli}. In this way, one would obtain:
$\b^{\r} = 9 $ $ \b^{f} = 21.5$ 
$\a_{\r}(0) = 0.57$, $  \a_{f}(0) = 0.43  $, $ \a' = 0.93\,\rm GeV^{-2}$.
The difference in the residues ($\b^{\r}\not= \b^{f} $) is largely
compensated by the different intercept and the  final results, as
expressed by \Eqn{s3} and \Eqn{s1}, would be unaffected
(changes would be less than $3\%$).

These results should hold, at least qualitatively, also if we enlarge
the basis of eigenstates to enforce the unitarity of the $S$-matrix.
To show it explicitly implies a considerable amount of work, but we can be
convinced of it by going to the approximation of neglecting the pion
exchange contribution $\p$  in \Eqn{amp3} and \Eqn{amp1} because of its
smallness, and introducing, similarly to \cite{Zheng} and \cite{Donoghue}, two
effective states $|3>$ and $|4>$ to take into account the inelastic
scatterings of $\bar{D}^{*}\p$ and $\bar{D}^{*}\r$ states
respectively. In this way the unitary $4\times 4$ $S$-matrix can be 
constructed

\starteq
S^{(0)I} = \pmatrix{\eta e^{2i\d} & 0 & 
i\sqrt{1 - \eta^{2}}e^{i(\d + \d_{1} )} & 0 \cr
0 &  \eta e^{2i\d} & 0 & i\sqrt{1 - \eta^{2}}e^{i(\d + \d_{2})} \cr 
i\sqrt{1 - \eta^{2}}e^{i(\d + \d_{1} )} & 0 & \eta e^{2i\d_{1}} & 0 \cr
0 & i\sqrt{1 - \eta^{2}}e^{i(\d + \d_{2})}& 0 & \eta e^{2i\d_{2}}  
}  \nonumber
\bigskip
\fineq
where, for $I=3/2$, $\h = 0.76$ and $\d =- 0.04$, while for
$I=1/2$, $\h = 0.71$ and $\d = -0.02$ ($\d_{1}$ and $\d_{2}$ are free
parameters in this model). For reasonable values of $\d_{1}$, $\d_{2}$
($ \d_{i} \approx \d$), we basically find the same result as in \cite{Zheng},
i.e. that, within an uncertainty of $10-20\%$, the results that are
obtained by including final state interactions do not differ from those
obtained by the bare amplitudes with no final state interaction at all.

We can apply these results to the calculation of
the width of some two-body $B$ decays into charmed
states. With inelastic effects given mostly
 by the Pomeron contribution and neglecting the small strong phases
$\d$, the full decay
amplitudes $A$ for $B \rarr D^{*}\p$ and $B \rarr D\r$, according to 
\Eqn{amp}, are (see also \cite{Zheng}):
\starteq
A = xA_{b} + \sqrt{1 - x^{2}}A'_{b}
\label{fullamp} 
\fineq
where $A'_{b} $ is the $B$ decay amplitude into an inelastic final state
and $x = \sqrt{(1+\eta)/2}.$

The decay rates with inelastic effects included can be obtained from
\starteq
|A|^{2} = |A_{b}|^{2} + (1 - x^{2})(|A'_{b}|^{2} - |A_{b}|^{2})~.
\label{fullrate} 
\fineq
Now, if $A'_{b} $ is comparable to $A_{b} $, the decay rate thus
obtained would be close to the rate obtained without final state 
interaction effects. This agrees well with experiment 
\cite{Neubert,Kamal,hon} for neutral $B$ decays into charged final
states (e.g. $D^{-}\p^{+}$, $D^{-}\r^{+}$, $D^{*-}\p^{+}$) where the
factorization model works rather well for those amplitudes which do not
receive contributions from the short-distance operator $O_{2}$ which
depends on $a_{2}$. For those amplitudes which depend on both 
the short-distance operators $O_{1}$ and $O_{2}$, such as  $B^{+}$
decays into $\bar{D}^{0}\p^{+}$, $\bar{D}^{0}\r^{+}$, $\bar{D}^{*0}\p^{+}$,
the rates obtained in the factorization model without final state 
interaction effects are significantly smaller than the experimental
results  using the value
$a_{2} \simeq 0.11$ as predicted by QCD. This suggests that 
 in $B$ decays there might be
non-factorization contributions \cite{Blok,Cheng}.

  Which conclusions can we draw from our analysis? As a preliminary
remark, let us 
observe that, even though we have dealt here with specific decay
channels, some aspects of our analysis are general enough to be applied
also to other decay modes. For example we have found that the inelastic
amplitude $ D^{*}\p \rarr D \r$ is strongly suppressed as compared to
the elastic ones. This result does not depend on the dominance,
in this case, of the
pion exchange which has a small intercept, i.e. $\a_{\p}(0) \approx 0$,
but would hold quite generally because of the suppression
of the non-leading Regge trajectories as compared to the
Pomeron contribution. Therefore the only surviving inelastic effects
at $\sqrt s = m_B$ should consist of
inelastic multiparticle production and should
be dominated by Pomeron exchange. We 
have found, in agreement with \cite{Zheng}, that
also these inelastic effects should not destroy the predictions for 
exclusive two-body non-leptonic $B$ decays based on factorization and
perturbative QCD. In \cite{Donoghue}, charmless final states were
considered and the opposite conclusion was reached. The apparent
contrast has its origin in the hypothesis
contained  in \Eqn{betaPcu}, that
strongly reduces the Pomeron coupling to charm and therefore reduces
the inelastic effects in all the channels with charmed mesons in the
final state. This explains why \cite{Donoghue} finds significant final
strong interaction effects in the considered channels.
\Eqn{betaPcu} has been assumed
in \cite{Zheng}  on a pure theoretical ground, since there is
no experimental information on $\beta^P(cu)$. However
a dependence on the quark mass can
be clearly seen in the $K p$ total cross section which is asymptotically
smaller than in  $\p p$ scattering~; a fit to the asymptotic cross
section is obtained by taking a reduction of $2/3$ in the Pomeron-quark
residue: $\b^P (su) \simeq {2 \over 3}\b^P (uu)$; this shows
a rather strong decrease with the quark mass. 
A powerlike dependence on the quark mass is certainly compatible with 
the result of \Eqn{betaPcu}, and, 
therefore, even if a clear numerical extrapolation is
difficult, the assumption given by \Eqn{betaPcu}
is reasonable. Therefore we are led to the conclusion
that final state interactions produce only moderate effects in the 
amplitudes provided that there are charmed (with or without open charm)
particles among the decay products. On the other hand,
for light charmless particles in the final state, the
Pomeron contribution increases and rescattering effects may alter
significantly the simple predictions based on perturbative QCD and heavy
quark effective theory.
 
\bigskip

\startbib

\bibitem{Buccella} F. Buccella, M. Forte, G. Miele and G. Ricciardi, 
\zph{48} (1990) 47.

\bibitem{Lusignoli} F. Buccella, M. Lusignoli and A. Pugliese, {\em Phys. 
Lett.} {\bf B 379} (1996) 249.
 
\bibitem{Despande} N. G. Despande and C. O. Dib, \PL{319} (1995) 313.

\bibitem{Zheng} H. Zheng, \PL{356} (1995) 107.

\bibitem{Donoghue} J. F. Donoghue, E. Golowich, A. A. Petrov and
J. M. Soares, {\em Phys. Rev. Lett.} {\bf 77} (1996) 2178.

\bibitem{CP} A. J. Buras and M. K. Harlander, in
{\em Heavy Flavors}, eds.  A. J. Buras and H. Lindner,  World
Scientific, Singapore (1992).

\bibitem{Bj} 
J. D. Bjorken, {\em Nucl. Phys.} {\bf B} (Proc. Suppl.) {\bf 11} (1989)
325.
 
\bibitem{Neubert} M. Neubert, V. Rieckert, B. Stech and Q. P. Xu, in
{\em Heavy Flavors}, eds.  A. J. Buras and H. Lindner,  World
Scientific, Singapore (1992).

\bibitem{Kamal} A. N. Kamal and T. N. Pham, \PRD{50} (1994) 395.

\bibitem{Collins} See for example, P. D. B. Collins, in {\em Regge theory
and high energy physics}, Cambridge University Press, 1977. 

\bibitem{hon} T. E. Browder, K. Honscheid and D. Pedrini, {\em
Nonleptonic Decays and Lifetimes of $b$-quark and $c$-quark Hadrons},
hep-ph/9606354, to appear in {\em Annual Review of Nuclear and Particles
Science, Vol. 46}.

\bibitem{Watson} K. M. Watson, {\em Phys. Rev.} {\bf 88} (1952) 1163.

\bibitem{Donnachie} A. Donnachie and P. V. Landshoff, \PL{296} (1992) 227.

\bibitem{PDG} Particle Data Group, R. M. Barnett {\em et al.}, 
\PRD{54}(1996) 1.

\bibitem{Irving} A. C. Irving and R. P. Worden, {\em Phys. Rep.}
 {\bf 34} (1977) 117.

\bibitem{Colangelo} P. Colangelo, A.Deandrea, N. Di Bartolomeo, 
F. Feruglio, R. Gatto and G. Nardulli,  \PL{339} (1994)151; 
the scaling law for $g_{B^{*}B\p}$   
was previously obtained by T.~N. Pham, \PRD{25} (1982) 2955. 

\bibitem{Deandrea} A.Deandrea, N. Di Bartolomeo, R. Gatto and    
G. Nardulli, \PL{318} (1993) 549.

\bibitem{Giacomelli} For a review, see, e.g.,
G. Giacomelli , {\em Phys. Rep.} {\bf 23} (1976) 123.

\bibitem{Blok} B. Blok and M. Shifman, \NP{389} (1993) 534. 

\bibitem{Cheng} H. Y. Cheng  , \PL{335} (1994) 428.

\endbib

\end{document}